# Subject grounding to reduce electromagnetic interference for MRI scanners operating in unshielded environments

Beatrice Lena[1], Bart de Vos[1], Teresa Guallart-Naval[2], Javad Parsa[1], Pablo Garcia Cristobal[2], Ruben van den Broek[1], Chloé Najac[1], Joseba Alonso[2], Andrew Webb[1]

**Affiliations:**

[1] *C.J. Gorter MRI Center, Leiden University Medical Center, Leiden, Netherlands*

[2] *Instituto de Instrumentación para Imagen Molecular (I3M), Centro Mixto CSIC-Universitat Politècnica de València, 46022 Valencia, Spain*

**Corresponding author:** Beatrice Lena, Leiden University Medical Center, Albinusdreef 2, 2333 ZG Leiden. Email: b.lena@lumc.nl

**Funding:** Dutch Science Foundation Open Technology 18981. Ministerio de Ciencia e Innovación of Spain (PID2022-142719OB-C22). European Innovation Council (EIC-Transition 101136407). Horizon 2020 ERC Advanced Grant (670629).

# Abstract

**Purpose:**

Portable low-field (<0.1 T) MRI is increasingly used for point-of-care imaging, but electromagnetic interference (EMI) presents a significant challenge, especially in unshielded environments. EMI can degrade image quality and compromise diagnostic utility. This study investigates whether subject grounding can effectively reduce EMI and improve image quality, comparing different grounding strategies.

**Methods:**

Experiments were conducted using a 47 mT Halbach-based MRI scanner with a single receive channel. Reproducibility was evaluated at a second site using a 72 mT scanner with similar geometry. Turbo spin echo sequences were used to image the hand and brain. Subject grounding was implemented using conductive cloth sleeves or ECG electrodes, each connected between the subject's skin and scanner ground. Three EMI conditions were tested: ambient, added single-frequency EMI, and broadband EMI. Signal-to-noise ratios (SNR) were calculated under each configuration.

**Results:**

Subject grounding significantly reduced EMI in both hand and brain scans. In hand imaging, conductive sleeves reduced noise from 85× to 1.25× the 50-ohm noise floor. In brain imaging, grounding alone reduced noise from 55× to 25× baseline; when combined with arc RF shields, noise was further reduced to 1.2× baseline, even under added EMI. These results were reproducible across different scanners and locations.

**Conclusion:**

Subject grounding is a simple, effective, and reproducible strategy for mitigating EMI in portable low-field MRI. It is especially effective for hand imaging, while brain imaging benefits from additional RF shielding. The approach is robust under various EMI conditions and may complement other denoising techniques.

# 1. Introduction

Recent advances in system hardware architecture and deep-learning (DL)-based image processing algorithms have led to a renewed interest in the design of low-field (≤ 0.1 T) MRI systems within point-of-care (POC) settings [1-3]. One of its biggest challenges is electromagnetic interference (EMI). Unlike conventional MRI, portable low-field systems operate without a Faraday cage, making them susceptible to external EMI, especially in environments such as intensive care units (ICUs) and emergency rooms, where medical devices such as ventilators, infusion pumps, and patient monitors can degrade image quality, sometimes rendering images completely non-diagnostic.

At the low frequencies (~1-4 MHz) at which POC systems operate, the human body acts as an efficient monopole antenna, coupling EMI into the receiver coil [4]. A number of potential solutions have been proposed to address this issue. One approach involves the use of conductive cloth around the subject [5, 6]. While very simple and effective, the performance of conductive cloths has been seen to degrade over time.

Most approaches are implemented in post-processing by utilising data from external sensors, such as additional RF coils [7-9], together with either analytically- or DL-based algorithms for EMI-suppression. The EDITER method (External Dynamic InTerference Estimation and Removal), for instance, uses external interference detection and template subtraction to remove EMI in an analytic manner before reconstruction [7]. More recently, deep learning-based methods [8, 10-12] have shown strong performance in reducing the EMI contribution. Although these approaches have been demonstrated to be highly effective, an inherent drawback of the multiple external detector approach is the increased system complexity and the necessity for a multi-channel receiver, in addition to the potential for location-dependent performance, due to the training of the DL algorithm and the positioning of the sensors.

An alternative approach is the use of lightweight flexible RF shields placed around the subject [13]. This approach results in a reduction in broadband environmental-based EMI levels that is almost equivalent to that with no subject present. However, if there are strong localized sources of EMI in the vicinity, this approach may not give sufficient suppression. Moreover, in some cases such as patients in the ICU, RF shielding may not always be practical.

In this study, we assess an alternative/complementary approach to mitigate EMI using subject grounding. This approach uses either standard ECG electrodes or simple conductive pads, which connect the subject directly to ground. A series of experiments was conducted at two sites across a range of scenarios in order to evaluate the efficacy of the proposed grounding technique and intersite-reproducibility in reducing EMI.

# 2. Methods
## 2.1 MR acquisition

All data at the Leiden University Medical Center (site 1) were acquired using a 47 mT (1.98 MHz) Halbach-based MRI scanner [14] with a Magritek Kea2 spectrometer (Aachen, Germany) with a single receive channel and a built-in transmit/receive switch and preamplifier. For hand imaging, a cylindrical saddle transmit coil combined with an ellipsoidal solenoid receive coil was used. The transmit coil had a 19 cm diameter, was 27 cm long, and had 3 turns of 1.5 mm diameter litz wire, while the solenoid coil had a 13 cm width, 11 cm height, was 19 cm long, and had 20 turns of 1.5 mm litz wire. The Q of the receive coil was 400 and 440, for the loaded and unloaded case, respectively, while the Q of the transmit coil was 70 in both cases. Brain images were acquired using a semi-ellipsoidal solenoid transmit/receive coil with a 18 cm width, and 24 cm height, 16 cm long, and had 15 turns of 1.5 mm diameter litz wire. The Q of this coil was 160 and 220, for the loaded and unloaded case, respectively.

Turbo spin echo (TSE) sequences were used to acquire images in vivo, with the following hardware setups and MR parameters:

- *Hand:* TR/TE/TE$_{eff}$ 400/16/16 ms, echo train length 8, Field-of-View (FOV) [RL x FH x AP] 160 × 200 × 100 mm$^3$, center-out Cartesian k-space trajectory, imaging bandwidth 20 kHz, number of data points 160 × 200 × 33, voxel size 1 × 1 × 3 mm$^3$, total acquisition time 4:40 minutes.
- *Brain:* TTR/TE/TE$_{eff}$ 500/19/19 ms, echo train length 6, FOV [AP x RL x FH] 240 × 210 × 200 mm$^3$, center-out Cartesian k-space trajectory, imaging bandwidth 20 kHz, number of data points 160 × 140 × 40, voxel size 1.5 × 1.5 × 5 mm$^3$, total acquisition time 07:45 minutes.

Informed consent was obtained and this study has been approved by the local ethics committee (NL83272.058.22) under the stipulation that the maximum scan time for a subject is limited to 60 minutes.

## 2.2 Experiments

A sleeve made of conductive cloth (4711 series, Holland Shielding Systems BV, Dordrecht, the Netherlands) was used to ground the subject for hand imaging, while two ECG electrodes (3M Red Dot ECG electrodes), connected in parallel and placed on the arm and lower leg, were used to ground the subject for the neuroimaging scans (see Figure S1). The sleeve and ECG electrodes were connected to the front panel of the magnet. For the brain scans two semi-cylindrical aluminium RF shields (each 1.5 m wide and 1 m long) were added for a subset of experiments.

Images were acquired with and without grounding the subject, and compared to the noise floor as measured by a 50 Ω load, for the following three scenarios:

i) Environmental EMI,
ii) Adding a discrete frequency EMI via a transmitting antenna and function generator. The antenna was positioned outside the RF shields, transmitting a 1.981 MHz continuous sine wave with 8 mV and 20 mV peak-to-peak voltages for imaging of the brain and hand, respectively.
iii) Adding 2 MHz broadband EMI with the same setup as ii) with 1 V peak-to-peak amplitude for both brain and hand images.

The noise levels including EMI were quantified by acquiring a noise profile (50 kHz bandwidth) for each of the aforementioned scenarios. The noise level was calculated from a TSE sequence in which no RF pulse was applied, with 1,000 repetitions. The discrete Fourier transform of these measurements gave 1,000 frequency-dependent noise sensitivity profiles, from which the standard deviation as a function of frequency was calculated [9]. A polynomial function was then fitted to the noise profile (NP), in order to remove the effects of the coil frequency profile: the maximum value was used to assess the noise level (indicated in the paper as $\sigma(NP)$). Additionally, the image signal-to-noise ratio (SNR) was quantified and calculated as the ratio of the mean signal measured in a region of interest (ROI) in the centre of the brain/wrist, with a relatively uniform image intensity, and the standard deviation of the noise measured from four ROIs in the corners of the image [15]. For cases in which the SNR was expected to be very low, the noise distribution deviated from Gaussian and followed a Rician distribution. In such cases, the SNR was estimated using the adapted formula:

$$\text{SNR} = \frac{\text{mean(signal)}}{\sqrt{2} \times \text{std (noise)}}$$

to account for the statistical properties of the noise distribution in the low-SNR regime [16].

In addition, to test the repeatability of the measurements, the SNR was measured in brain scans of two subjects over four days, with and without subject grounding. To determine whether there were significant differences in SNR values between the two conditions, a two-sample t-test was conducted with significance level α of 5%.

## 2.3 Reproducibility of the method

As part of the 2024 ISMRM reproducibility challenge, researchers from the Instituto de Instrumentación para Imagen Molecular in Valencia independently replicated this study using a 72 mT Halbach-based MRI scanner [5]. The system was controlled via MaRCoS, an open-source, high-performance Magnetic Resonance Control System [17, 18]. Wrist imaging was performed on a healthy volunteer after obtaining written informed consent in accordance with institutional guidelines. Images were acquired with a single-channel solenoidal transmit/receive coil, using a turbo spin echo sequence with the following parameters: TR/TE/TE$_{eff}$ 200/10/10 ms, echo train length 10, Field-of-View (FOV) [RL x FH x AP] 180 × 140 × 80 mm$^3$, center-out Cartesian k-space trajectory, imaging bandwidth 20 kHz, number of data points 80 × 80 × 10, voxel size 2.25 × 1.75 × 8 mm$^3$, total acquisition time 6:40 minutes. In each instance, noise scans were obtained, from which the root-mean-square (RMS) value of each acquisition was calculated. Noise was measured with different subject grounding approaches and compared against a reference measured with a 50-ohm load: various numbers of ECG electrodes (same brand as in Leiden) were used as well as copper tape and a conductive sleeve. The same experiments were repeated in Leiden to check the reproducibility of the results by calculating the experimental noise factor (NF), defined as:

$$\text{NF} = \frac{\text{Noise}_{\text{measured}}}{\text{Noise}_{\text{50 Ohm Load}}}$$

# 3. Results

Figure 1 showed that the EMI was comparable to the values of the noise measured with a 50 Ω load connected to the spectrometer, both for the hand and the head scans, when grounded using conductive sleeve and two ECG electrodes, respectively. The 50 Ω load profile, representing the noise floor, showed a roll-off at the frequency extremes due to the effects of the digital filter[19].

In the hand scans (Figure 1a.), in the ungrounded condition the measured noise reached 85× the noise floor defined by the 50-ohm load. Subject grounding reduced this to a factor of 1.25, effectively restoring the system's intrinsic noise level. Similar results were observed when additional broadband EMI (92×baseline) or single-frequency EMI (75×baseline) was introduced. In all cases, subject grounding substantially mitigated EMI, keeping levels within twice the baseline.

For the brain scans (Figure 1b.), subject grounding reduced noise from approximately 55× the baseline to around 25× baseline. To further reduce this, arc shields were added in combination with grounding (Figure 1c). In the presence of environmental noise, the ungrounded configuration showed noise levels of 1.8× baseline, increasing to 4× baseline in the presence of intentionally introduced EMI. With grounding and arc shielding, noise was reduced to 1.2× baseline, approaching the noise floor. This combined approach yielded robust noise suppression, over a 3-fold reduction under EMI, highlighting its effectiveness even in challenging environments. Detailed quantitative values both for hand and brain scanning can be found in Supporting Information S2.

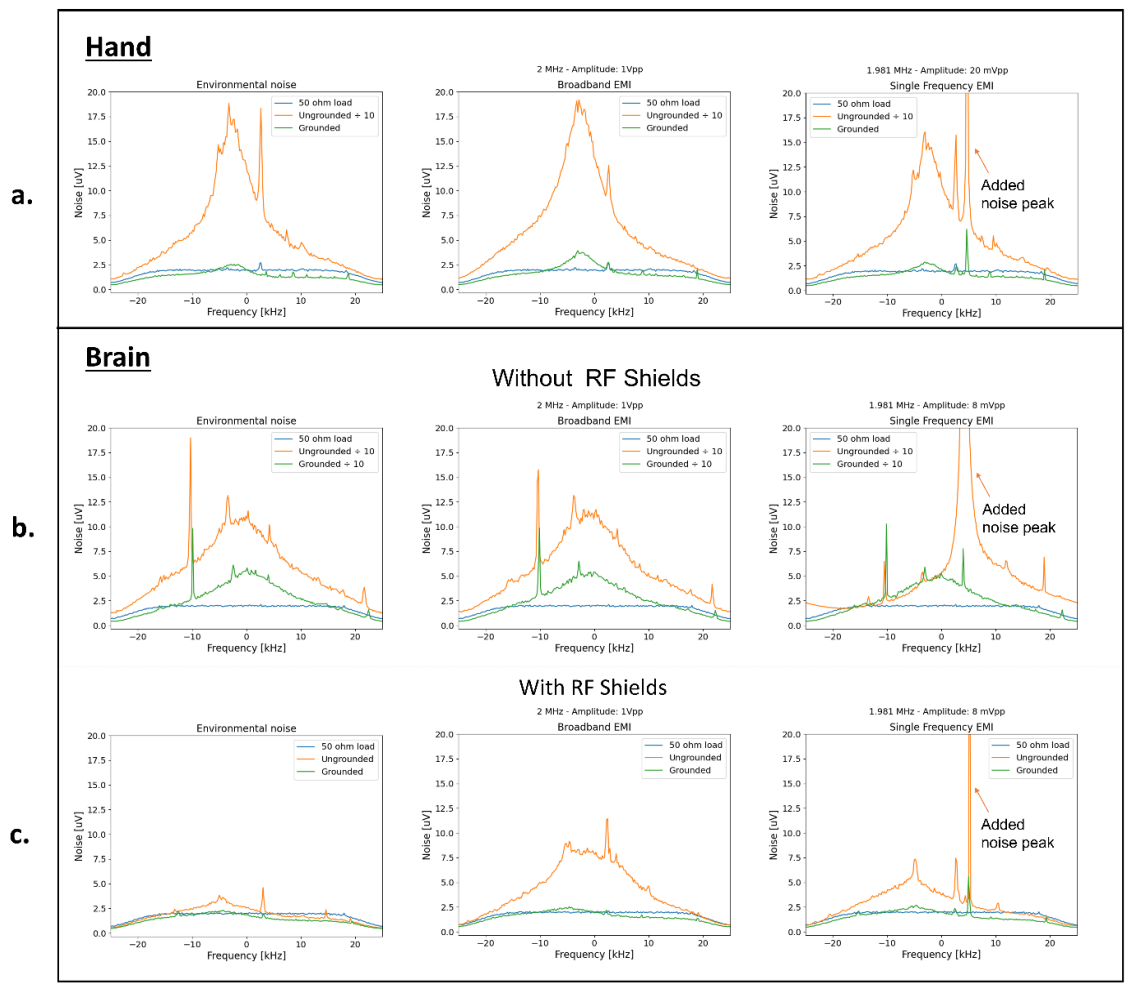

**Figure 1.** Noise levels measured for hand and brain scans as a function of the frequency, with and without grounding, for the different noise conditions described in the methods section. For the head scans, the cases with and without two additional RF shields were added. The reference noise level was obtained by connecting a 50 Ω load to the spectrometer. When not specified, the noise spikes visible at different frequencies were not added and are the contributions of unknown noise sources. In the plots on the right, the intentionally added EMI peaks had a maximum value of 400 for the hand and 1320 and 40, for the head without and with arc shields, respectively.

Figure 2a. shows TSE images of the hand: the environmental noise levels without subject grounding made the images unreadable, even when noise was not intentionally added. When broadband EMI was added, grounding the subject increased the SNR by a factor of ~4. With the single frequency EMI was applied, the SNR improved by a factor of ~2, and the discrete "zipper artefact" is less evident (see Table 1).

Figure 2b. illustrates the quality of the brain images acquired under different shielding configurations. Notably, when only subject grounding was applied, noise levels remained approximately 20 times above the noise floor. The addition of the two arc shields was essential to achieve cleaner images. When these shields were in place, subject grounding further improved image stability across all tested configurations, compared to using arc shields alone.

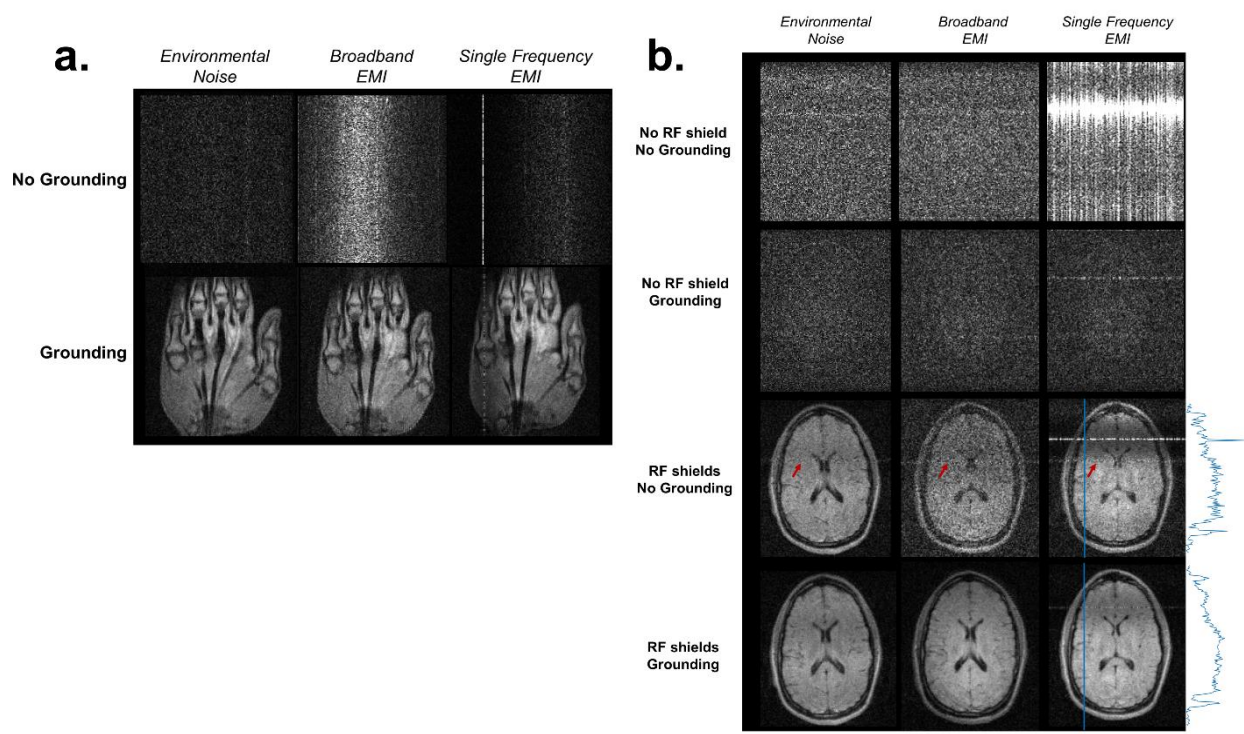

**Figure 2.** a. Hand images with and without the added EMI, with and without subject grounding. The use of RF shields for wrist imaging was not possible, the EMI during our typical scanning conditions exceeded the threshold for visibility. However, subject grounding enabled the visualization of the wrist with acceptable image quality, even in the presence of intentionally added EMI. b. Brain images with and without the added EMI, with and without subject grounding, in presence of two additional RF shields. The image quality clearly improved with subject grounding. On the case where single frequency noise was added and the arc shields were used, the 1D- profile of the signal across the blue line is also reported on the side. A zipper artefact, coming from an unknown noise source in the hospital, can be identified in the images without subject grounding (red arrows).

A summary of the average signal in a ROI, noise as standard deviations in 4 ROIs and SNR measured in hand and brain images can be found in Table 1. As expected, in all conditions with subject grounding, the SNR improved.

**Table 1.** Mean Signal, standard deviation of the noise and signal-to-noise ratio calculated for the brain and hand images in the conditions reported in Figure 2 and 3. For wrist imaging, where imaging was only possible using subject grounding, it was observed that even in the presence of added EMI, the SNR remained above 10. For brain imaging, the results presented include the use of arc shields. When arc shields were combined with subject grounding, the SNR was maximized, yielding the highest measured values. In the "Ungrounded" case for the hand imaging, no signal could be detected and measured due to the extremely high noise levels.

| | Body part | Environmental Noise | | Broadband EMI | | Single Frequency EMI | |
|---|---|---|---|---|---|---|---|
| Mean Signal | | Ungrounded | Grounded | Ungrounded | Grounded | Ungrounded | Grounded |
| | Hand | - | 1299 | - | 1269 | - | 1235 |
| | Brain | 1020 | 1015 | 1237 | 1176 | 1018 | 1043 |

| | Body part | Environmental Noise | | Broadband EMI | | Single Frequency EMI | |
|---|---|---|---|---|---|---|---|
| Std Noise | | Unground | Grounded | Ungrounded | Grounded | Ungrounded | Ground |
| | Hand | 9699 | 73 | 12021 | 79 | 3010 | 70 |
| | Brain | 72 | 49 | 149 | 49 | 106 | 49 |

| | Body part | Environmental Noise | | Broadband EMI | | Single Frequency EMI | |
|---|---|---|---|---|---|---|---|
| Signal to Noise Ratio | | Ungrounded | Grounded | Ungrounded | Grounded | Ungrounded | Grounded |
| | Hand | < 1 | 18 | < 1 | 16 | < 1 | 18 |
| | Brain | 14 | 20 | 8 | 24 | 10 | 21 |

In all five repeated instances of scanning the hands of the same two volunteers the noise was reduced, and the SNR correspondingly improved, when using grounding (Figure 3). The results of the t-test indicated a significant difference between the two conditions (t = 42.5, p < 0.01).

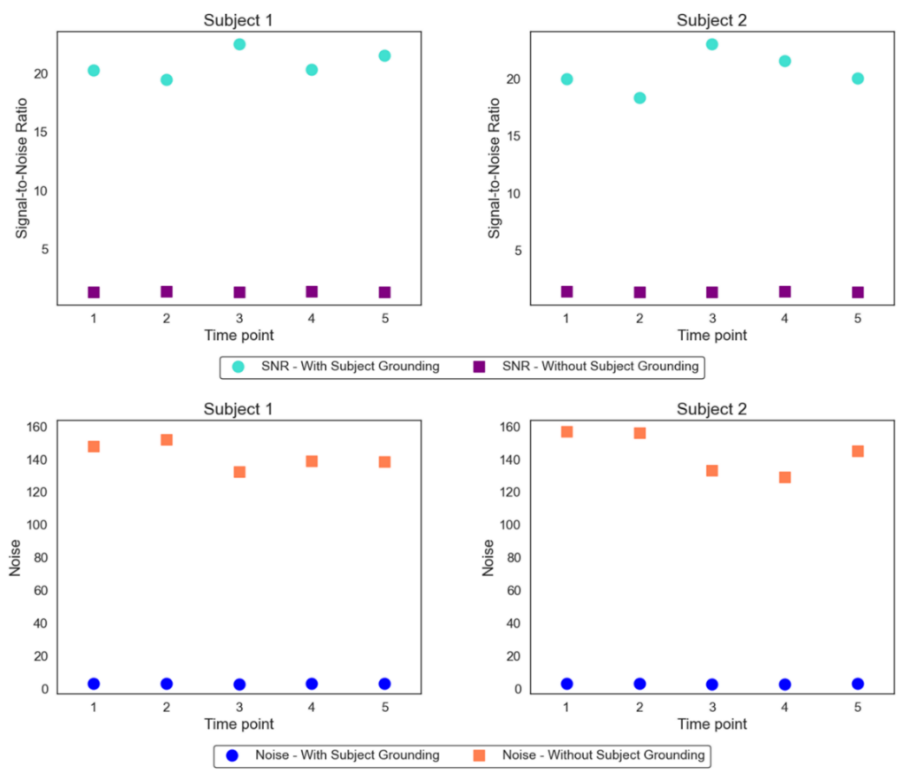

**Figure 3.** SNR and noise level measured from hand scans in 5 days for two subjects, with and without subject grounding. In all cases, subject grounding reduced the noise level and increased the SNR.

Figure 4 presents the noise factors and images obtained from acquisitions conducted in Valencia and in Leiden with various subject grounding techniques. The noise level due to the environment in Leiden was higher than in Valencia ("no grounding" case), but subject grounding methods were effective at reducing noise in both locations. Notably, the noise level decreased as the number of ECG electrodes increased. Alternative grounding methods proved to be more practical and effective in reducing noise. For instance, the use of a copper tape resulted in a noise level comparable to that of a 50-ohm load. A conductive fabric sleeve also demonstrated high effectiveness, while being more patient-friendly. Noise levels measured in all cases in both locations can be found in Supporting Information S3.

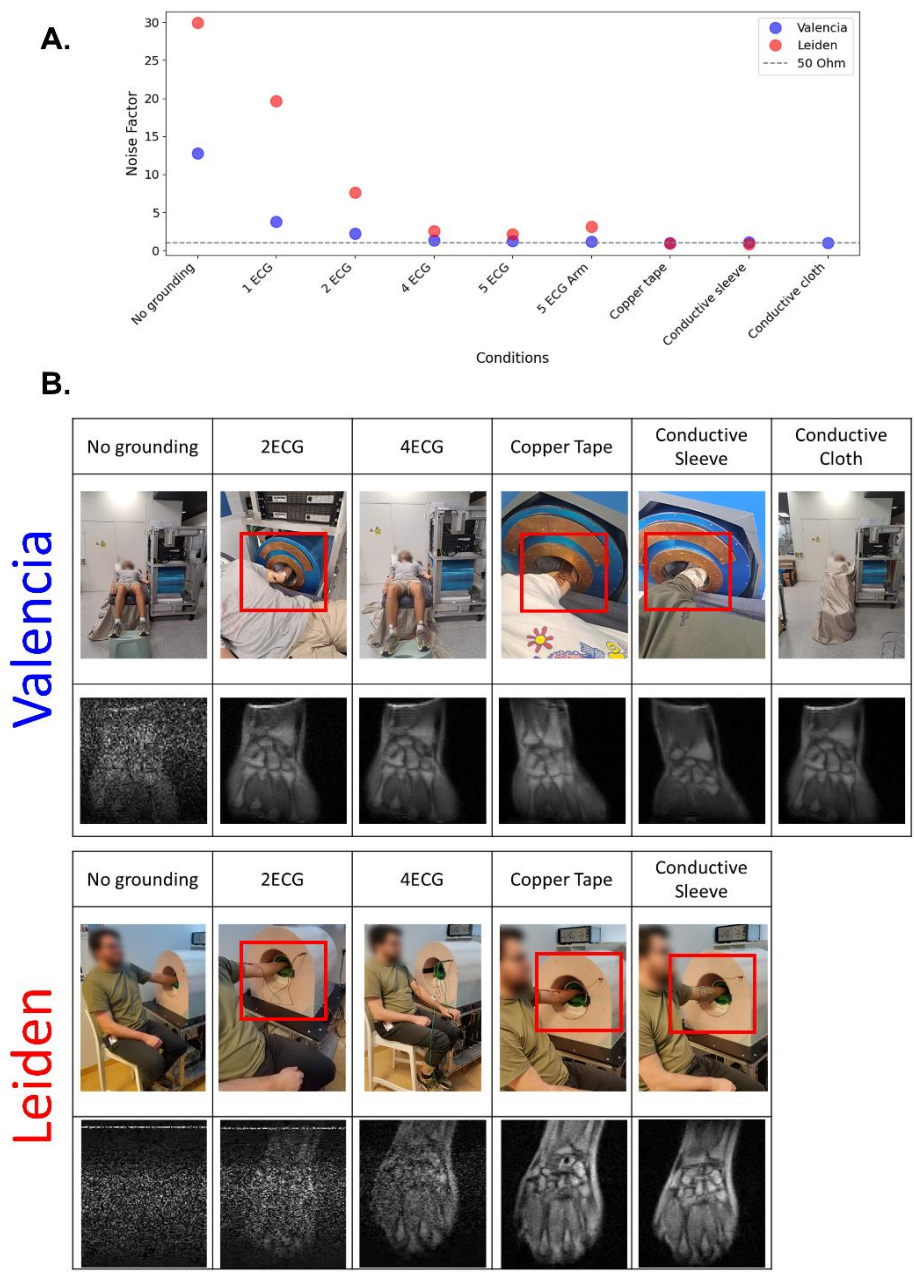

**Figure 4**. A. Noise factor vs different methods of achieving subject grounding, compared to the measured 50-ohm load (dashed grey line) for both measurements in Valencia and Leiden B. Positioning and example scan images for imaging the hand of a subject using different grounding techniques (in red squares). NOTE: The conductive cloth case could not be acquired in Leiden due to deterioration.

# 4. Discussion

In this study, we evaluate a simple approach for reducing EMI during scanning in unshielded environments with low-field MRI. This approach is straightforward and allows for imaging in areas where otherwise this would not be possible. In the case of brain imaging, the combination of subject grounding with flexible RF shields led to a statistically significant reduction in EMI and enhancement in SNR. In the case of hand imaging, where the use of flexible and removable RF shielding is not feasible, the application of a conductive sleeve for subject grounding also resulted in significant levels of EMI reduction. The results were shown to be reproducible for different volunteers and different locations, and are robust with respect to the nature of the EMI e.g. single frequency, broadband or ambient. This method is an effective means of reducing baseline noise, resulting in a more stable measurement setup, which can be used in different environments.

Some previous studies have used related approaches in which conductive materials are used as an RF shield and are connected to ground [5-7, 20]. In their paper [20], Nakagomi et al. proposed the use of a mobile 0.2 T MR scanner to study elbow injuries in baseball players. A conductive cloth was attached to the RF shield box to provide electromagnetic shielding for the subject and additional cloths were wound around the upper and lower arm of the subjects, to obtain high-quality images. As the cloths were attached to the RF box and wound around the subject's bare skin, it is probable that subject grounding was achieved without the researchers specifically noting this. Therefore, their findings are consistent with our own results. Similarly, Srinivas et al. [7] uses one ECG electrode attached to the wrist of the subjects, as EMI detector. Finally, the group in Valencia has also used a conductive cloth [5], wrapped around the subject and then grounded, as a method to improve the SNR.

The study by Kibret et al. [23] models the human body as a monopole antenna above a ground plane, in the context of wireless signal transmission around the body (body area networks). While their work was not focused on MRI, we found that this antenna-based framework offers a useful perspective for understanding our own observations in low-field MRI. Specifically, we observed that grounding the subject closer to the imaging region — for example, at the neck in neuroimaging or near the wrist for hand imaging — consistently reduced image noise (see Supporting Information S4). This may be due to a shorter effective antenna length, which reduces the strength of the associated electric field at the hands or top of the head, and its coupling with environmental noise. Conversely, grounding farther from the coil (e.g., at the hip or calf) increases the effective antenna length and appears to raise the noise level. These trends are supported by noise measurements provided in the supplementary material.

We note that subject grounding a standard procedure applied in a number of medical procedures (ECG, EMG, EEG). One electrode, the "ground electrode", is part of the Driven Right Leg circuit [21], meaning that it is not connected to any lead and has the specific purpose of minimizing the noise. The same principle is applied when using different types of subject grounding: the conductive sleeve, for example, acts as 6 or more ECG.

# 5. Conclusion

The subject grounding method effectively reduces external EMI in low-field MRI, enhancing SNR and image quality in unshielded environments. This approach is versatile, reproducible, and can be integrated with other denoising techniques, offering significant potential for use in various POC settings. The method can be employed in conjunction with other denoising techniques, such as EDITER [7], AUTOMAP [22], other AI-based methods [23, 24] or the noise removal achieved through noise sensors/single-coil-based method [25].

# 6. Acknowledgements

We would like to thank Tom O'Reilly for the implementation of the noise quantification script and sequence and the volunteers that have participated to this study. This work was funded by the Dutch Science Foundation Open Technology 18981, Ministerio de Ciencia e Innovación of Spain (PID2022-142719OB-C22), European Innovation Council (EIC-Transition 101136407). Horizon 2020 ERC Advanced Grant (670629).

# 8. Supporting Information

## S1. Scanning setup for hand and brain imaging

Subject grounding equipment in Leiden: the conductive sleeve for hand and ECG electrodes brain imaging.

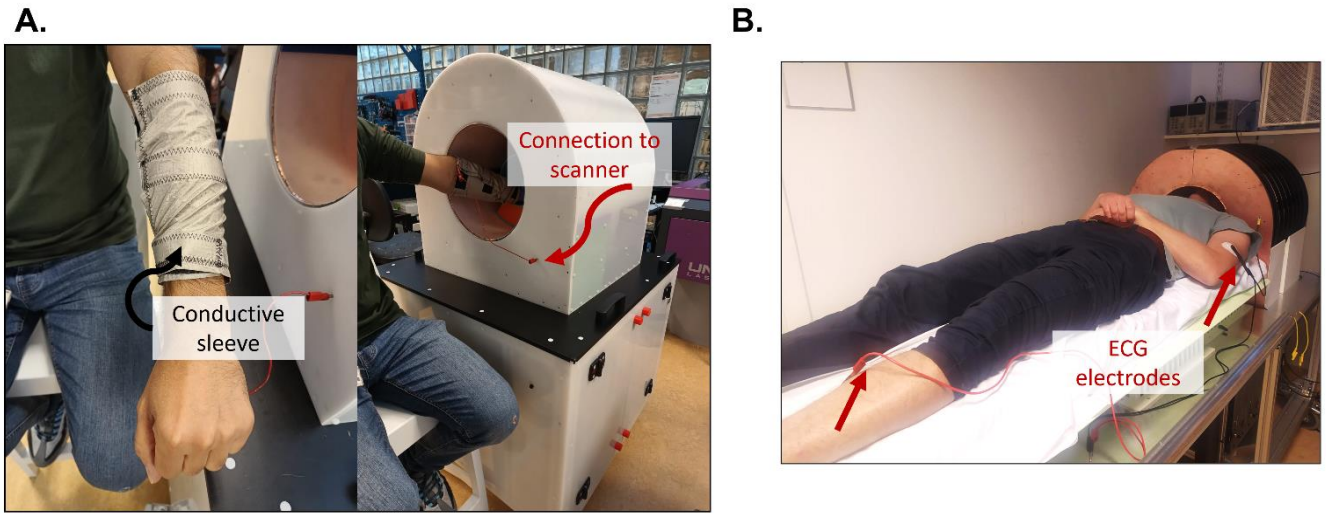

**Figure S1**. A. Conductive sleeve (black arrow) wrapped around a volunteer forearm and connected to the scanner (red arrow) for hand imaging; B. ECG electrodes (red arrow) connected to the arm and lower leg of the volunteer for brain imaging.

## S2. Noise Measurements in hand and brain, with and without subject grounding

Noise as measured from a TSE sequence in which no RF pulse was applied, through the standard deviation of 1,000 frequency-dependent noise sensitivity profiles ($\sigma(NP)$).

**Table S1**: Summary of the noise measurements in hand and brain with and without intentionally added EMI.

**Hand**

| Conditions | Environment $\sigma(NP)$ | Broadband $\sigma(NP)$ | Single Freq. $\sigma(NP)$ |
|---|---|---|---|
| No grounding | 170.3 | 184.7 | 149.1 |
| Grounding | 2.5 | 3.7 | 2.8 |
| 50-Ohm load | | 2.0 | |

**Brain**

*Without RF Shields*

| Conditions | Environment $\sigma(NP)$ | Broadband $\sigma(NP)$ | Single Freq. $\sigma(NP)$ |
|---|---|---|---|
| No grounding | 109.4 | 112.6 | 137.1 |
| Grounding | 54.5 | 52.3 | 48.2 |
| 50-Ohm load | | 2.0 | |

*With RF Shields*

| Conditions | Environment $\sigma(NP)$ | Broadband $\sigma(NP)$ | Single Freq. $\sigma(NP)$ |
|---|---|---|---|
| No grounding | 3.4 | 8.6 | 6.0 |
| Grounding | 2.2 | 2.4 | 2.5 |
| 50-Ohm load | | 2.0 | |

## S3. Noise Measurements in Leiden and Valencia

Noise as measured in the real part of the RMS of the noise measurements, in uV, and noise factors in Valencia and Leiden for different subject grounding methods.

Table S2: Summary of the noise measurements from each site.

| Conditions | Valencia | | Leiden | |
|---|---|---|---|---|
| | Noise [uV] | Noise Factor | Noise [uV] | Noise Factor |
| No grounding | 581.4 | 12.78 | 2396.38 | 29.95 |
| 1 ECG | 172.7 | 3.80 | 1569.42 | 19.62 |
| 2 ECG | 100.1 | 2.20 | 607.39 | 7.59 |
| 4 ECG | 60.6 | 1.33 | 203.57 | 2.54 |
| 5 ECG | 58.4 | 1.28 | 174.64 | 2.18 |
| 5 ECG Arm | 51.7 | 1.14 | 249.16 | 3.11 |
| Copper tape | 45.3 | 1.00 | 70.58 | 0.88 |
| Conductive Sleeve | 47.9 | 1.05 | 67.94 | 0.85 |
| Conductive Cloth | 44.0 | 0.97 | - | - |
| 50 Ohm | 45.5 | 1.00 | 80.01 | 1.00 |

## S4. Noise Measurements in different body parts

Noise as measured in the real part of the RMS of the noise measurements, in uV, and effective antenna length for two subjects, where the grounding was performed with copper tape, placed on the neck, waist and calf of the subjects. For comparison, the noise floor as measured by a 50 Ω load was 80 ± 2 uV.

Table S3: Summary of the noise measurements and antenna lengths in two subjects.

| Conditions | Subject 1 Noise [uV] | Subject 2 Noise [uV] | Conditions | Subject 1 Antenna Length [cm] | Subject 2 Antenna Length [cm] |
|---|---|---|---|---|---|
| No grounding | 2513 ± 80 | 1555 ± 93 | No grounding | 192 | 160 |
| Tape on calf | 743 ± 56 | 748 ± 35 | Tape on calf | 157 | 134 |
| Tape on waist | 200 ± 11 | 140 ± 4 | Tape on waist | 80 | 67 |
| Tape on neck | 69 ± 4 | 96 ± 2 | Tape on neck | 30 | 30 |

For comparison, the noise floor as measured by a 50 Ω load was 80 ± 2 uV.